\documentclass[11pt,a4paper]{article}

\newcommand{\cpfversion}{Version 0.9.6}
\newcommand{\cpfbuild}{20191025}

\usepackage[utf8]{inputenc}
\usepackage{amsmath}
\usepackage{amsfonts}
\usepackage{amssymb}
\usepackage{harpoon}	
\usepackage{graphicx}
\usepackage{color, colortbl}

\usepackage{fancyhdr}
\usepackage[backend=bibtex]{biblatex}   
\bibliography{papers}

\usepackage[nottoc]{tocbibind}

\usepackage[hyperindex = true, colorlinks = true, linktocpage = true, linkcolor=blue, citecolor=blue, bookmarks, pdftitle={Stochastic Algorithmic Differentiation of (Expectations of) Discontinuous Functions (Indicator Functions)}, pdfauthor={Christian Fries}]{hyperref}

\newcounter{cpf_counter} 
\newcounter{cpfNumberOfFigures} 
\newcounter{cpfNumberOfTables} 

\newcommand{\diag}{\mathrm{diag}}

\newcommand{\abs}[1]{{\vert #1 \vert}}

\newcommand{\indicatorfcn}{\mathrm{\mathbf{1}}}

\newcommand{\transposed}{\top}

\title{Stochastic Algorithmic Differentiation of (Expectations of) Discontinuous Functions (Indicator Functions)}

\author{%
	Christian P.~Fries\\
	{\small \href{mailto:email@christian-fries.de}{email@christian-fries.de}}\\
}

\date{November 14, 2018 \\ (this version: October 25, 2019)}

\begin{document}

\pagestyle{fancy}               

\DeclareGraphicsExtensions{.pdf,.jpg,.png}


\maketitle

\centerline{\small\href{http://www.christianfries.com/finmath/stochasticautodiff}{\small\cpfversion}}


\section*{Abstract}
	
	In this paper, we present a method for the accurate estimation of the derivative (aka.~sensitivity) of expectations of functions involving an indicator function by combining a stochastic algorithmic differentiation and a regression.
	The method is an improvement of the approach presented in \cite{FriesAutoDiff4MonteCarlo, FriesAutoDiff4AmericanMonteCarlo}. 
	
	The finite difference approximation of a partial derivative of a Monte-Carlo integral of a discontinuous function is known to exhibit a high Monte-Carlo error. The issue is evident since the Monte-Carlo approximation of a discontinuous function is just a finite sum of discontinuous functions and as such, not even differentiable. 

	The algorithmic differentiation of a discontinuous function is problematic. A natural approach is to replace the discontinuity by continuous functions. This is equivalent to replacing a path-wise automatic differentiation by a (local) finite difference approximation.
	
    We present an improvement (in terms of variance reduction) by decoupling the integration of the Dirac delta and the remaining conditional expectation and estimating the two parts by separate regressions. For the algorithmic differentiation, we derive an operator that can be injected seamlessly - with minimal code changes - into the algorithm resulting in the exact result. 
\vfill

\begin{footnotesize}
	
	\noindent \textbf{Disclaimer:} The views expressed in this work are the personal views of the authors and do not necessarily reflect the views or policies of current or previous employers. \newline
	Feedback welcomed at \href{mailto:email@christian-fries.de}{email@christian-fries.de}.
	
\end{footnotesize}


%

\newpage


\tableofcontents


\newpage

\section{Introduction}
	\label{sec:autodiffforindicators:introduction}

	We consider the numerical valuation of
	\begin{equation}
		\label{eq:autodiffforindicators:diffOfExpOfDisc}
		\frac{\partial}{\partial \theta} \mathrm{E} \left(
			f\left( \indicatorfcn_{X > 0}  \right)
		\right) \text{,}
	\end{equation}
	where $X$ denotes a random variable depending on an arbitrary model parameter $\theta$ , $\indicatorfcn_{X > 0}$ denotes the indicator function on the random variable $X > 0$, $f$ denotes a general operator on random variables and $\mathrm{E}$ is the expectation operator.

	If the expectation operator $\mathrm{E}$ is approximated by a Monte-Carlo integration, the path-wise differentiation of the discontinuity is problematic, see below. We investigate the performance of the method presented here using a Monte-Carlo integration, but the method is not specific for Monte-Carlo applications.
	
	Expressions like~\eqref{eq:autodiffforindicators:diffOfExpOfDisc} are common in computational finance, e.g.~in the valuation of digital options. We choose this example as a test case to discuss some numerical results in Section~\ref{sec:autodiffforindicators:numericalResults}. However, the numerical procedure presented below is general in purpose.
	
	\subsection{Derivatives of Monte-Carlo Integrals of Discontinuous Functions}

	The numerical calculation of partial derivatives of valuations obtained by Monte-Carlo integration of discontinuous functions is a well-known problem. Different numerical approaches have been considered. The path-wise analytic differentiation results in a distribution, which sometimes can be integrated analytically, allowing to replace the differentiation of the discontinuous function by an analytic solution, examples are~\cite{JoshiKainth2004DeltasOfNthToDefaultSwaps, RottFries2005GreeksForCDOs}. The path-wise finite-difference approximation is known to result in substantial Monte-Carlo errors (for small finite-difference shifts) or biased derivatives (for large finite-difference shifts), cf.~\cite{GlassermanPaul2003, FriesLectureNotes2007}.
	
	An alternative approach consists of the differentiation of the probability density, resulting in the likelihood ratio method \cite{GlassermanPaul2003} or more general Malliavin calculus \cite{Benhamoud2003Malliavin}. Proxy simulation schemes provide a methodology to obtain likelihood ratio sensitivities seamlessly by finite differences, see~\cite{FriesKampen2007ProxyScheme}. However, the likelihood ratio method produces undesired large Monte-Carlo variance for Monte-Carlo integrals of smooth functions, \cite{FriesLectureNotes2007}.
	
	\sloppypar Partial proxy simulation schemes \cite{FriesJoshi2008PartialProxyScheme} and localized proxy simulation schemes \cite{Fries2007LocalizedProxyScheme} combine the two approaches to reduce the Monte-Carlo variance and optimal localizers can be derived, see~\cite{ChanJoshi2015OptimalLimitProxyScheme}, which also contains an excellent review of the topic. However, such schemes are specific to the nature of the random variables entering the indicator function (speaking of valuation of financial products, they are product (or pay-off) dependent), i.e.~they require to set up a specific simulation for each valuation. This results in a degradation of the performance of the numerical algorithm with respect to CPU time. In application like the valuation of portfolio effects (like xVAs), such a product dependent simulation is unfeasible.
	
	\subsection{Algorithmic Differentiation and Stochastic Algorithmic Differentiation}

	Algorithmic differentiation \cite{Griewank2008} propagates the analytic partial derivative through a valuation algorithm to obtain a numerical value of the derivative, which can approximate the analytic solution up to a numerical error due to the propagation of the floating-point errors of each individual operation. A derivative obtained from an algorithmic differentiation is highly accurate as long as the involved functions are differentiable and their derivative is known with high accuracy.
	
	For the Monte-Carlo simulation, a classic application of algorithmic differentiation results in a highly accurate path-wise differentiation. Discontinuous functions are an issue, and a common approach is to replace the discontinuous function by a smoothed approximation, then differentiate the approximation.
	If the smoothing is a linear function, it gives results identical to a classical path-wise finite-difference. This approach leaves us with the problem to find the appropriate smoothing function or interval.
	
	\medskip
	
	A stochastic algorithmic differentiation is an algorithmic differentiation (formally) applied to random variables, see~\cite{FriesAutoDiff4MonteCarlo}. It thus allows inspecting the stochastic nature of the random variables and their derivatives, e.g., the variance of the argument of the indicator function.
	
	As \textit{expected stochastic algorithmic differentiation}, we denote the stochastic algorithmic differentiation where the outer (last) operator is an expectation (or conditional expectation). This allows to explore the stochastic nature of the differentials further. See~\cite{FriesAutoDiff4AmericanMonteCarlo, FriesAutoDiff4MonteCarloForwardSensitivities} for examples.

	\section{Derivative of the Indicator as Conditional Expectation}

	We consider a stochastic algorithmic differentiation, see~\cite{FriesAutoDiff4MonteCarlo}, where the algorithmic differentiation is applied to random variables, and the final operation is an expectation. Let $X$ denote a random variable depending on an arbitrary model parameter $\theta$ and $\indicatorfcn_{X > 0}$ the indicator function on the random variable $X > 0$ and $f$ a general operator on random variables. Then we consider
	\begin{align*}
		\frac{\partial}{\partial \theta} \mathrm{E} \left(
			f\left( \indicatorfcn_{X > 0}  \right)
		\right)
		& \ = \ 
		\mathrm{E} \left(
			f^{\prime} \circ \frac{\partial}{\partial \theta} \indicatorfcn_{X > 0}
		\right) \\
		& \ = \ 
		\mathrm{E} \left(
			f^{\prime} \circ \frac{\partial}{\partial X} \indicatorfcn_{X > 0} \circ \frac{\partial X}{\partial \theta}
		\right)
		\ = \ 
		\mathrm{E} \left(
			f^{\prime} \circ \delta(X) \circ \frac{\partial X}{\partial \theta}
		\right) \text{,}
	\end{align*}
	where $\delta$ is the Dirac delta interpreted as a linear operator.

	Note that it is sufficient to consider an expression of this form. For the general case of a function $f$ depending on multiple-arguments, the algorithmic differentiation will decompose the differentials into sums of such expressions and expressions not involving indicator functions, which are not in our interest.
	
	\medskip
	
	
	
	
	Assume that $\theta$ is a scalar model parameter. The general case will follow by interpreting the following result component-wise for every component of the parameter $\theta$. If $\theta$ is a scalar, we can rewrite
	\begin{equation*}
		\frac{\partial X}{\partial \theta} \ = \ \diag\left({\frac{\partial X}{\partial \theta}}\right) \circ \indicatorfcn \text{,}
	\end{equation*}
	where $\diag$ is just the corresponding linear operator (a diagonal matrix if we consider a finite sample space) and $\indicatorfcn$ is the random variable being identical to $1$.
	
	If we interpret the expectation operator as a scalar-product, that is
	\begin{equation*}
		\mathrm{E} \left(
			Z
		\right)
		\ := \ 
		\langle \indicatorfcn  \, , Z \rangle \ := \ \int_{\Omega} \indicatorfcn(\omega) \cdot Z(\omega) \ \mathrm{d}P(\omega) \text{,}
	\end{equation*}
	we can write
	\begin{align*}
		\mathrm{E} \left(
			f^{\prime} \circ \delta(X) \circ \frac{\partial X}{\partial \theta}
		\right)
		& \ = \
		\mathrm{E} \left(
			f^{\prime} \circ \delta(X) \circ \diag\left({\frac{\partial X}{\partial \theta}}\right) \circ \indicatorfcn
		\right) \\
		& \ = \
		\langle \indicatorfcn  \, , f^{\prime} \circ \delta(X) \circ \diag\left({\frac{\partial X}{\partial \theta}}\right) \circ \indicatorfcn \rangle \\
		& \ = \
		\langle A  \, , \delta(X) \circ \diag\left({\frac{\partial X}{\partial \theta}}\right) \circ \indicatorfcn \rangle \text{,}
	\end{align*}
	where $A \ := \ \left( \indicatorfcn^{\transposed} \circ f^{\prime} \right)^{\transposed}$ is the adjoint derivative.\footnote{In an adjoint algorithmic differentiation, the random variable $A$ is the derivative vector that is propagated backward though the chain rule.}

	With this definition we have
	\begin{equation*}
		\mathrm{E} \left(
			f^{\prime} \circ \delta(X) \circ \frac{\partial X}{\partial \theta}
		\right)
		\ = \
		\mathrm{E} \left(
			A \cdot \delta(X) \cdot \frac{\partial X}{\partial \theta}
		\right) \text{,}
	\end{equation*}
	where the multiplications $\cdot$ are now path-wise multiplications of random variables.\footnote{With a slight abuse of notation we do not make a difference between the linear operator $\delta(X)$ and the random variable $\delta(X) \indicatorfcn$, both acting as Dirac delta.}	
	
	Furthermore we find
	\begin{align}
		\notag
		\mathrm{E} \left(
			A \cdot \delta(X) \cdot \frac{\partial X}{\partial \theta}
		\right)
		& \ = \
		\mathrm{E} \left(
			A \cdot \frac{\partial X}{\partial \theta} \cdot \delta(X)
		\right) \\
		\notag
		& \ = \
		\mathrm{E} \left(
			\mathrm{E} \left(
				A \cdot \frac{\partial X}{\partial \theta} \ \vert \ X
			\right)
			\cdot \delta(X)
		\right) \\
		\label{eq:stochaadforindicator:densitydecomp}
		& \ = \
		\mathrm{E} \left(
			A \cdot \frac{\partial X}{\partial \theta} \ \vert \ X = 0
		\right)
		\cdot \phi_{X}(0) \text{,}		
	\end{align}
	where $\phi$ is the density function of $X$ and $\delta$ is the Dirac delta function.\footnote{
		It is $\mathrm{E} \left( \mathrm{E} \left( Z \ \vert \ X \right) \cdot \delta(X) \right) = \int_{-\infty}^{\infty} \mathrm{E} \left( Z \ \vert \ X=x \right) \delta(x) \phi_{X}(x) \mathrm{d} x = \mathrm{E} \left( Z \ \vert \ {X = 0} \right) \cdot \phi_{X}(0)$.
	} 
	The decomposition \eqref{eq:stochaadforindicator:densitydecomp} is important for us, since it enables us to improve the numerical approximation of the derivative by using different methods for the approximation of the conditional expectation and the density. 
	
	
	\subsection{Backward (Adjoint) Derivative}
	\label{sec:stochaadforindicator:backwardDifferentationAsConditionalExpectation}
	
	We like to derive a decomposition like~\eqref{eq:stochaadforindicator:densitydecomp} for the case of a backward (adjoint) algorithmic differentiation. The challenge in the algorithmic differentiation is to find an expression that we can apply the moment the adjoint derivative operator $A \frac{\mathrm{d}}{\mathrm{d} \theta}$ hits the indicator function $\indicatorfcn_{X>0}$, because this will greatly simplify the implementation.

	Assume that $\mathrm{E} \left( \indicatorfcn_{X=0} \right) > 0$. Then we have for the conditional expectation
	\begin{equation*}
			\mathrm{E} \left(
				A \cdot \frac{\partial X}{\partial \theta} \ \vert \ X = 0
			\right)
			\ = \ 
			\mathrm{E} \left(
				A \cdot \frac{\partial X}{\partial \theta} \cdot \frac{\indicatorfcn_{X=0}}{\mathrm{E} \left( \indicatorfcn_{X=0} \right)}
			\right)
			\ = \ 
			\mathrm{E} \left(
				A \cdot \frac{\indicatorfcn_{X=0}}{\mathrm{E} \left( \indicatorfcn_{X=0} \right)} \cdot \frac{\partial X}{\partial \theta}
			\right) \text{.}
	\end{equation*}
	Thus we find with \eqref{eq:stochaadforindicator:densitydecomp}
	\begin{equation}
		\label{eq:stochaadforindicator:exactFormulaWithDensity}
		\mathrm{E} \left(
			A \cdot \frac{\partial}{\partial \theta} \indicatorfcn_{X > 0} \cdot \frac{\partial X}{\partial \theta}
		\right)
		\ = \
		\mathrm{E} \left(
			A \cdot \frac{\indicatorfcn_{X=0}}{\mathrm{E} \left( \indicatorfcn_{X=0} \right)} \phi_{X}(0) \cdot \frac{\partial X}{\partial \theta}
		\right) \text{.}
	\end{equation}
	In other words, in an adjoint algorithmic differentiation, we may replace the derivative of the indicator function $\frac{\partial}{\partial \theta} \indicatorfcn_{X > 0}$ by $\frac{\indicatorfcn_{X=0}}{\mathrm{E} \left( \indicatorfcn_{X=0} \right)} \phi_{X}(0)$ and - in expectation - the result will agree with the true solution.
	
	For the case where $\{ X = 0 \}$ is a null-set, we assume that
	\begin{equation*}
		\lim_{w \rightarrow 0} \mathrm{E} \left( \ \cdot \ \vert \ \indicatorfcn_{\abs{X} \leq w} \right) \ = \ \mathrm{E} \left( \ \cdot \ \vert \ \indicatorfcn_{X=0} \right)
	\end{equation*}
	and obtain for $w$ small
	\begin{equation}
		\label{eq:stochaadforindicator:approxFormulaWithDensity}
		\mathrm{E} \left(
			A \cdot \frac{\partial}{\partial \theta} \indicatorfcn_{X > 0} \cdot \frac{\partial X}{\partial \theta}
		\right)
		\ \approx \
		\mathrm{E} \left(
			A \cdot \frac{\indicatorfcn_{\abs{X}<w}}{\mathrm{E} \left( \indicatorfcn_{\abs{X}<w} \right)} \phi_{X}(0) \cdot \frac{\partial X}{\partial \theta}
		\right) \text{.}
	\end{equation}


	\section{Regression Approximation of the Density}
	\label{sec:stochaadforindicator:regressionOfDensity}
	
	For the expression \eqref{eq:stochaadforindicator:densitydecomp} we need to estimate a conditional expectation $\mathrm{E}( A\ \vert \ X = 0)$ and the density $\phi_{X}(0)$. For the expressions \eqref{eq:stochaadforindicator:exactFormulaWithDensity} and \eqref{eq:stochaadforindicator:approxFormulaWithDensity}  we need to estimate the density $\phi_{X}(0)$.

	\smallskip
	
	Since our algorithm can inspect the random variable $X$ (the argument of the indicator function), we may approximate $\phi_{X}(0)$ by a regression.

	Let $\mathbb{Q}$ denote the probability measure and $d(x) := \frac{1}{x} \mathbb{Q}\left( X \in [0,x] \right)$. Then $\phi_{X}(0) = \lim\limits_{x \rightarrow 0} d(x)$. We approximate $\mathbb{Q}\left( X \in [0,x] \right)$ by counting the number of sample paths $X(\omega)$ falling into the interval $[0,x]$ and define
	\begin{equation*}
		\tilde{d}(x) \ := \
		\frac{1}{x} \frac{\abs{ \omega_{i} \ \vert \ X(\omega_{i}) \in [0,x] , \, i=0, \ldots,n-1}}{n} \qquad \text{for\ } x \neq 0 \text{.}
	\end{equation*}
	We then approximate $\tilde{d}$ by a local linear regression $d^{*}$ of the sample points $(X(\omega_{i}), \tilde{d}(X(\omega_{i})))$ for paths $\omega_{i}$ where $0< \abs{X(\omega_{i})} \leq w/2$.\footnote{The point $x=0$ is excluded for obvious reasons.}

\begin{figure}[hbtp]
		\begin{center}
			\includegraphics[scale=0.7]{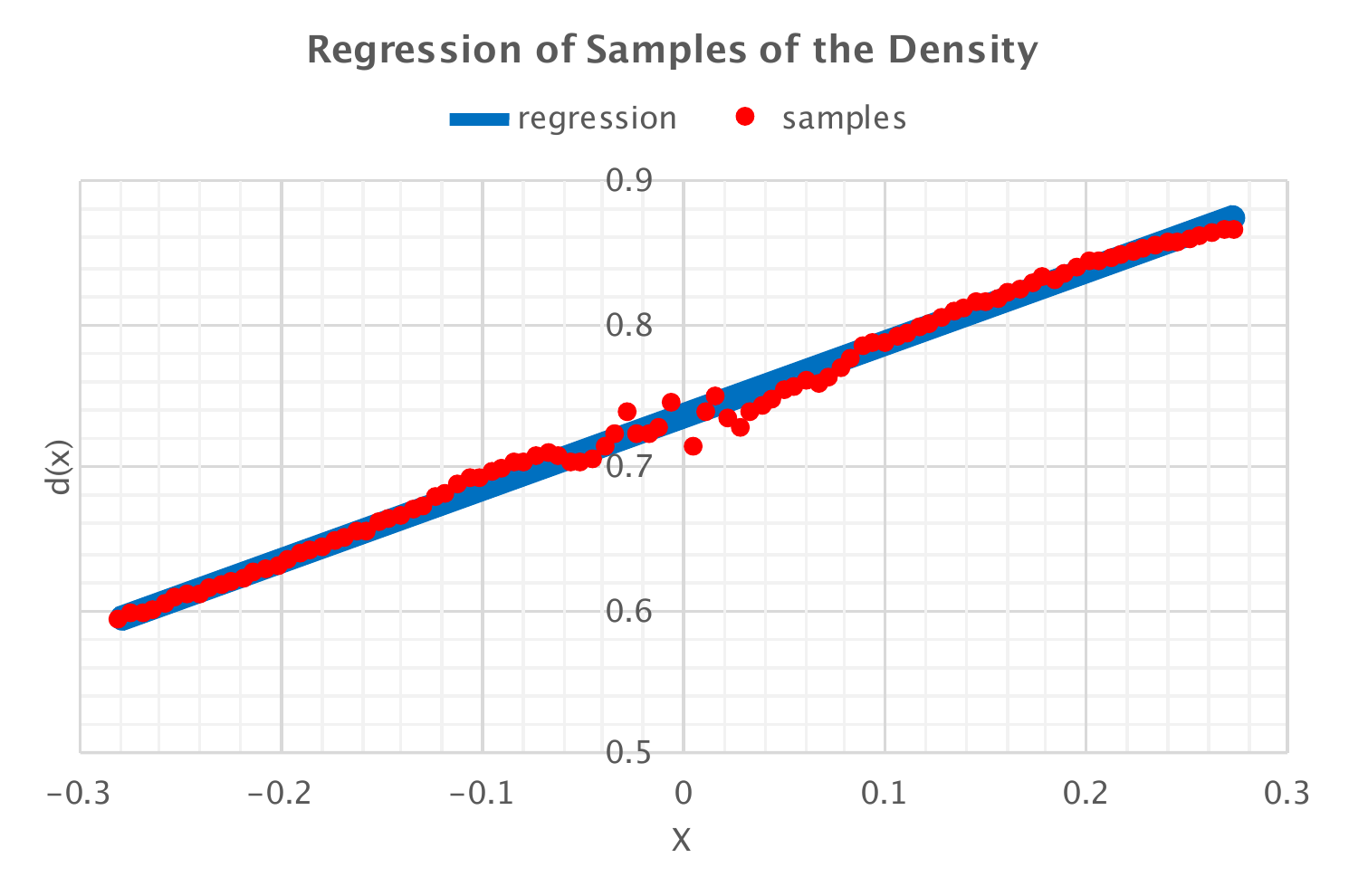}
			\caption[
			]{
				Regression of samples $(X(\omega_{i}), \tilde{d}(X(\omega_{i})))$ resulting in an approximation $\phi_{X}(0) \ \approx \ \tilde{d}(0)$.
			}
			\label{fig:RegressionOfSamplesOfDensity}
		\end{center}
		\addtocounter{cpfNumberOfFigures}{1}
	\end{figure}

	The width parameter $w$ can be chosen such that \textit{enough} sample points are used for the regression. In Figure~\ref{fig:RegressionOfSamplesOfDensity}, we show an example of the regression function $d^{*}$ of the samples of $(X(\omega_{i}), \tilde{d}(X(\omega_{i})))$ resulting in an approximation $\phi_{X}(0) \ \approx \ d^{*}(0)$ taken from our numerical experiments (see Section~\ref{sec:autodiffforindicators:numericalResults}).

	\paragraph{Non-linear Regression.}

	To localize the regression and remove the singularity $X = 0$, we define regression basis functions as multiple of $\indicatorfcn_{0 < \abs{X}< w/2}$, with some width parameter $w$.
	Using basis function $B_{i} = \indicatorfcn_{\abs{X}< w/2} X^{i}$ for $i=0,\ldots,m-1$ and defining the regression matrix $B = \left( B_{0}, \ldots, B_{m-1} \right)$ we get
	\begin{equation*}
		d^{*}(X) \ = \ B \left( B^{\transposed} B \right)^{-1} B^{\transposed} \tilde{d}(X)
	\end{equation*}
	and
	\begin{equation*}
		d^{*}(0) \ = \ \left(\indicatorfcn_{\abs{X}< w/2}, 0, \ldots, 0\right) \left( B^{\transposed} B \right)^{-1} B^{\transposed} \tilde{d}(X) \text{.}
	\end{equation*}
	
	\subsection{Regression of the Distribution Function}
	
	Alternatively, one may consider a regression of the distribution $D(x) := \mathbb{Q}\left( X \in [0,x] \right)$ with $D(0) = 0$ and take the linear regression coefficient as an approximation for $\phi_{X}(0)$.

	\subsection{Regression Approximation of the Conditional Expectation}

	For a backward algorithmic differentiation, we will use the expression~\eqref{eq:stochaadforindicator:exactFormulaWithDensity} or~\eqref{eq:stochaadforindicator:approxFormulaWithDensity}, which only require the estimation of $\phi_{X}(0)$. Here we shortly mention the estimation of the conditional expectation $\mathrm{E}\left( A \ \vert \ X=0 \right)$ in~\eqref{eq:stochaadforindicator:densitydecomp} by a standard linear regression. The derivation is helpful to understand why the decomposition~\eqref{eq:stochaadforindicator:densitydecomp} and te expressions~\eqref{eq:stochaadforindicator:exactFormulaWithDensity} and~\eqref{eq:stochaadforindicator:approxFormulaWithDensity} result in a significant improvement of the numerical accuracy.
	
	\medskip
	
	To localize the regression, we define basis function as multiple of $\indicatorfcn_{\abs{X}< w/2}$, with some width parameter $w$. Let $a^{*}(x)$ denote the regression approximation of $\mathrm{E}\left( A \ \vert \ X=x \right)$, then $a^{*}(0)$ is the corresponding approximation of $\mathrm{E}\left( A \ \vert \ X=0 \right)$.

	\paragraph{Non-linear Regression.}
	
	Using basis function $B_{i} = \indicatorfcn_{\abs{X}< w/2} X^{i}$ for $i=0,\ldots,m-1$ and defining the regression matrix $B = \left( B_{0}, \ldots, B_{m-1} \right)$ we get
	\begin{equation*}
		\mathrm{E}\left( A \ \vert \ X \right) \ \approx \ B \left( B^{\transposed} B \right)^{-1} B^{\transposed} A
	\end{equation*}
	and thus
	\begin{equation*}
		\mathrm{E}\left( A \ \vert \ \{ X = 0 \} \right) \ \approx \ \left(\indicatorfcn_{\abs{X}< w/2}, 0, \ldots, 0\right) \left( B^{\transposed} B \right)^{-1} B^{\transposed} A \text{.}
	\end{equation*}

	\paragraph{Linear Approximation.}

	For the special case $m = 2$ we have the explicit solution
	\begin{equation}
		\mathrm{E}\left( A \ \vert \ \{ X = 0 \} \right) \ \approx \ \frac{E(\tilde{X}^2) E(\tilde{A}) - E(\tilde{X}) E(\tilde{A} \tilde{X})}{E(\tilde{X}^2) - E(\tilde{X})^2} \text{,}
	\end{equation}
	where $\tilde{X} = \indicatorfcn_{\abs{X}< w/2}  X$ and $\tilde{A} = \indicatorfcn_{\abs{X}< w/2} A$.

	\paragraph{Projection.}
	
	For the special case $m = 1$ we find the projection approximation corresponding to~\eqref{eq:stochaadforindicator:approxFormulaWithDensity} as
	\begin{equation}
		\mathrm{E}\left( A \ \vert \ \{ X = 0 \} \right) \ \approx \ \frac{\mathrm{E}\left( \indicatorfcn_{\abs{X}< w/2} A \right)}{\mathrm{E}\left( \indicatorfcn_{\abs{X}< w/2} \right)} \text{.}
	\end{equation}

	\subsection{Recovering the Classic Discretized Delta Function Approximation}

	The classic approximation using a discretized delta distribution, which is also equivalent to a pay-off smoothing with a linear interpolation, is given by the approximation	
	\begin{equation}
		\label{eq:autodiffforindicators:classicalFDApproxOfDerivOfIndicator}
		\frac{\partial}{\partial X} \indicatorfcn_{X > 0} \ \approx \ \frac{1}{w}
		\begin{cases}
			1 & \text{for\ } \abs{X} < w/2 \\
			0 & \text{else.} \\
		\end{cases}
	\end{equation}
	This corresponds to a regression with a single basis function $\indicatorfcn_{\abs{X}< w/2}$ for the density and the conditional expectation ($m=1$). Hence, we recover this approach as being the simplest ($0$-order) approximation, having the (strong) constrain that both parts are estimates with the same basis function.
			
	\subsection{The Intuition behind the Approach}
	
	In Section~\ref{sec:autodiffforindicators:numericalResults} we show that the regression of the density results in a strong improvement with an almost optimal variance reduction. To understand why this is the case, let us compare the approach with the discretized delta function.
	Let
	\begin{equation*}
		\tilde{\indicatorfcn}_{w}(X) \ := \
		\begin{cases}
		1 & \text{for\ } \abs{X} < w/2 \\
		0 & \text{else.} \\
		\end{cases}
	\end{equation*}

	Then we have that
	\begin{equation}
		\label{eq:autodiffforindicators:approxOfA}
		\frac{1}{\mathrm{E}\left( \tilde{\indicatorfcn}_{w}(X) \right)} \mathrm{E}\left( A \cdot \tilde{\indicatorfcn}_{w}(X) \right)
	\end{equation}
	is an approximation of $E( A \ \vert \ X = 0 )$. If $A$ is a smooth function of $X$ at $X=0$, then the Monte-Carlo approximation of \eqref{eq:autodiffforindicators:approxOfA} does not suffer from increasing Monte-Carlo errors as $w$ tends to zero, as long as the Monte-Carlo simulation has at least one path in the set $\{ \abs{X} < w/2 \}$.
	
	Furthermore,
	\begin{equation}
		\label{eq:autodiffforindicators:approxOfPhi}
		\mathrm{E}\left( \frac{1}{w} \tilde{\indicatorfcn}_{w}(X)  \right)
	\end{equation}
	is an approximation of $\phi(x)$, but the Monte-Carlo approximation of \eqref{eq:autodiffforindicators:approxOfPhi} has a Monte-Carlo error which increases when $w$ tends to zero.
	
	The essence in our approach is to separate the two parts (which is possible) and use the fact that the stochastic algorithmic differentiation \cite{FriesAutoDiff4MonteCarlo} provides us with full knowledge of $X$. We may use a regression or a larger interval $w$ for the estimation of \eqref{eq:autodiffforindicators:approxOfPhi} and a smaller interval $w$ for the estimation of \eqref{eq:autodiffforindicators:approxOfA}.

	On the other hand, the classical approach \eqref{eq:autodiffforindicators:classicalFDApproxOfDerivOfIndicator} corresponds to
	\begin{equation}
		\label{eq:autodiffforindicators:approxOfAandPhi}
		\mathrm{E}\left( A \frac{1}{w} \tilde{\indicatorfcn}_{w}(X)  \right) \text{,}
	\end{equation}
	which will either generate a high Monte-Carlo variance for small $w$ (due to the density approximation) or a biased mean for large $w$ (due to the approximation of A).

	\section{Implementation}
	
	The random variable $X$ is the argument of the indicator function and can be extracted directly from the calculation by intercepting the implementation of the indicator function.
	
	\subsection{Modification of the Algorithmic Differentiation}
	
	Combining~\eqref{eq:stochaadforindicator:approxFormulaWithDensity} and Section~\ref{sec:stochaadforindicator:regressionOfDensity}
	we can inject our approximation into the algorithmic differentiation by using
	\begin{equation}
		\label{eq:autodiffforindicators:injectionProjectionAndDensity}
		\frac{\partial}{\partial X} \indicatorfcn_{X > 0} \ :\approx \ \frac{1}{\mathrm{E}\left( \indicatorfcn_{\abs{X} < w/2} \right)} \indicatorfcn_{\abs{X} < w/2} \ d^{*}(0) \text{,}
	\end{equation}
	where $d^{*}(0)$ is the regression approximation of $\phi_{X}(0)$. Given that the algorithm interprets operators as operators on random variables, this allows to implement the method by a small and local code change, just replacing the definition of $\frac{\partial}{\partial X} \indicatorfcn_{X > 0}$.
	
	A reference implementation can be found in \cite{finmath-lib} (version 3.6 or later).

	\subsection{Explicitly extracting the Jump Size and Speed}
	
	In case we like to analyse the random variable $A$ (which contains the jump size and the speed of the discontinuity), this can be done with a simple trick: we first run an automatic differentiation, where the differentiation of the indicator function is replaced by a unit $1$ and define the result as $A_{1}$. We then run the automatic differentiation again, where the differentiation of the indicator function is replaced by $0$ and define the result as $A_{0}$. We then have $A = A_{1} - A_{0}$.

	While this is not the most efficient way to extract $A$, it is a simple one, which may be used with a minimal modification of an existing algorithm.

	\newpage

	\section{Numerical Results}
	\label{sec:autodiffforindicators:numericalResults}

	\subsection{Test Setup}
	The advantage of the methodology can be analysed by considering the simple example of a digital option under a Black-Scholes model. We compare different approaches:
	\begin{itemize}
		\item A classical central finite difference approximation with different shift sizes.
		
		\item An expectation using an analytic likelihood ratio weight applied to the pay-off. Note that for the digital option, this approach is almost the optimal numerical method (apart from an additional importance sampling). However, the approach uses an analytic formula for the density, which is in general not available. Here, the method serves as a benchmark.

		\item An adjoint automatic differentiation (AAD) implementation (\cite{FriesAutoDiff4MonteCarlo}), where the differentiation of the indicator function is replaced by a \textit{discretized} delta function, i.e.,
		\begin{equation*}
			\frac{\partial}{\partial X} \indicatorfcn_{X > 0} \ \approx \ \frac{1}{w}
			\begin{cases}
				1 & \text{for\ } \abs{X} < w/2 \\
				0 & \text{else.} \\
			\end{cases}
		\end{equation*}
		This approach is effectively equivalent to a pay-off-smoothing, however, we can use the freedom to determine the width $w$ adaptively by inspecting properties of $X$. In the results the approach is labelled as \textit{Stoch.~AD}.

		\item The stochastic AAD implementation with an regression approximation of the density $\phi_{X}(0)$ using a linear regression using \eqref{eq:autodiffforindicators:injectionProjectionAndDensity}. In the results the approach is labelled as \textit{Stoch.~AD with Regression}.
	\end{itemize}
	Since the analytic solution is known, we may compare the numerical methods against the analytic solution.

	Our test setup is a Black-Scholes model with interest rate at $r=5\%$, volatility at $\sigma = 50\%$, initial value at $S_{0} = 1.0$. The digital option has a maturity of $T = 1.0$ and a strike of $K = 1.05$. The Monte-Carlo simulation uses 200000 paths with a Mersenne Twister pseudo random number generator.
	
	Within this setup we calculate the delta of the option ($\frac{\partial}{\partial S_{0}}$) using different seeds for the random number generator and compare the numerical results for the various approaches using different values for the shift parameter $w$. The shift parameter $w$ was used consistently for all methods (excepts for the analytic and likelihood ratio method, of course). We used~\eqref{eq:autodiffforindicators:injectionProjectionAndDensity} in the stochastic AAD with regression.
	
	\sloppypar The interval used for the regression of the density of $X$ was $\pm \frac{w_{\phi}}{2} \mathrm{stddev}(X)$ with $w_{\phi} = 0.5$ (Table~\ref{tbl:DeltaDistribution-Width-05}, Table~\ref{tbl:DeltaDistribution-Width-005}) and $w_{\phi} = 0.25$ (Table~\ref{tbl:DeltaDistribution-Width-0025-reg025}). The regression was a first order regressions in $X$ ($m=2$) on the distribution function.

	We repeat each Monte-Carlo experiment 10000 times and report the average of the Monte-Carlo values. This allows us to identify a possible bias in the Monte-Carlo estimator. Note that if a Monte-Carlo method exhibits a bias, then the result of 10000 Monte-Carlo experiments with 200000 paths is not the same as one experiment with 1000 million paths.
	
	The results presented here can be reproduced by the code in \cite{finmath-lib}\footnote{See \texttt{MonteCarloBlackScholesModelDigitalOptionAADRegressionSensitivitiesTest}.}
	
	\subsection{Test Results}
	
\begin{figure}[!h]
	\begin{center}
		\includegraphics[scale=0.7]{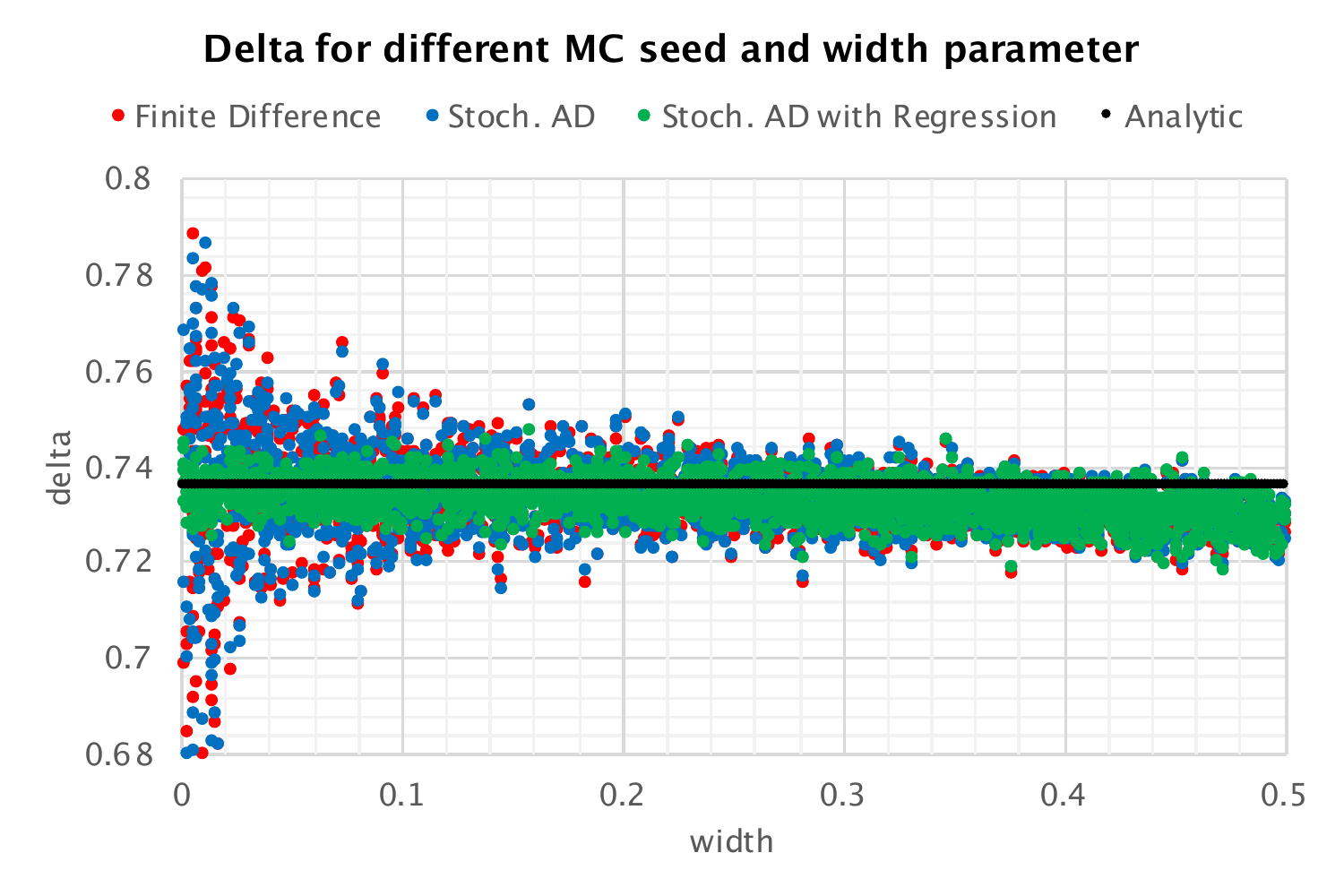}
		\caption[
		]{
			Delta values of digital option obtained from different Monte-Carlo simulations (seeds) and different width parameter.
		}
		\label{fig:DeltaForSeedAndWidth-FDvsAADvsReg}
	\end{center}
	\addtocounter{cpfNumberOfFigures}{1}
\end{figure}

	Figure~\ref{fig:DeltaForSeedAndWidth-FDvsAADvsReg} shows the delta of the digital option using different values for the width parameter $w$. The experiment is repeated for different Monte-Carlo seeds to assess the Monte-Carlo error. As expected, the finite difference method has a high Monte-Carlo error for small shift size. The algorithmic differentiation with discretized delta function (pay-off smoothing) shows the same behaviour as the classical finite difference: it also suffers from increasing Monte-Carlo variance as the width parameter $w$ becomes small. The algorithmic differentiation with density regression gives stable results, only suffering from a biased estimate for larger values of $w$.

\begin{figure}[hbtp]
	\begin{center}
		\includegraphics[scale=0.7]{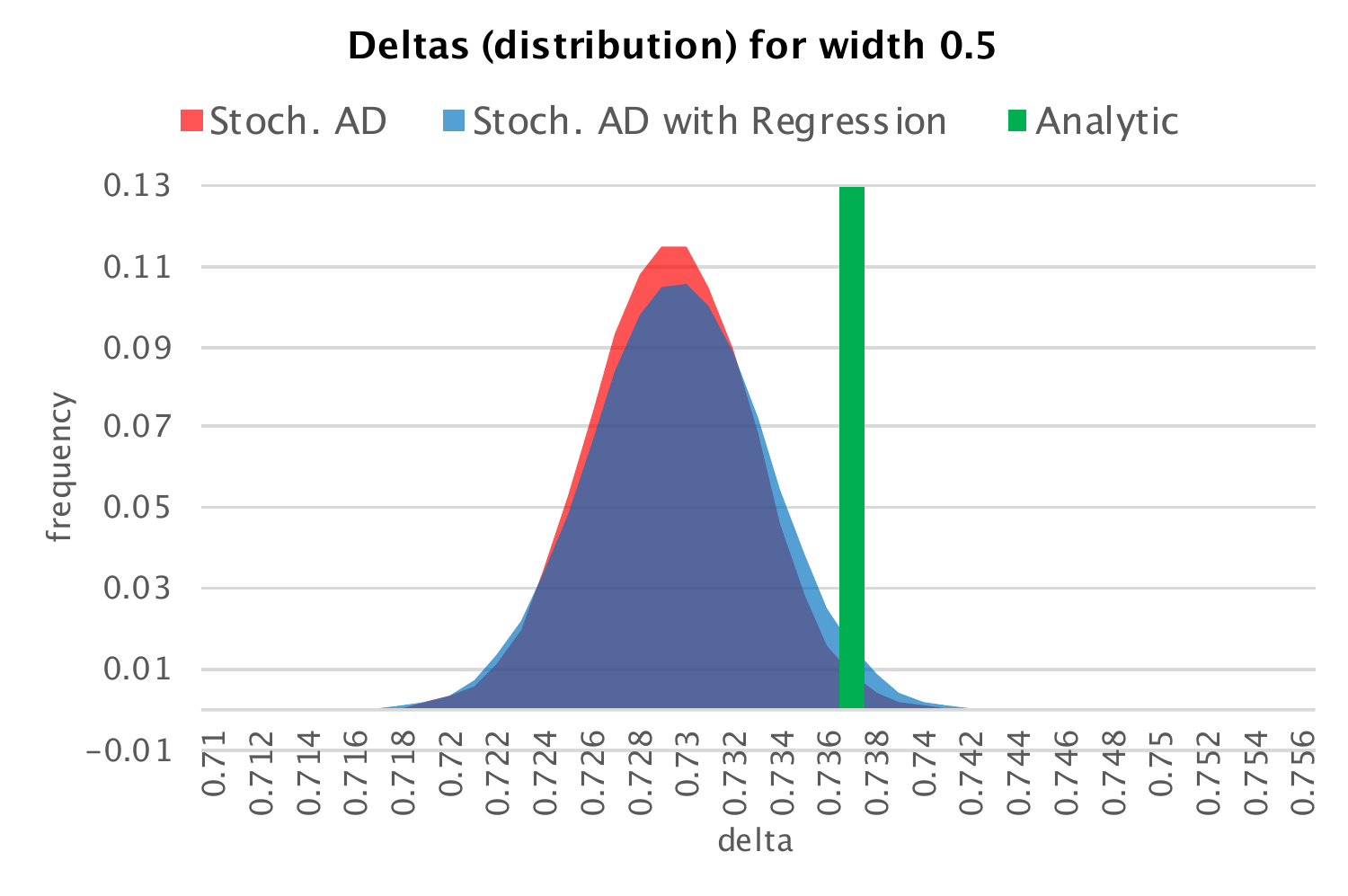}
		\caption[
		]{
			Distribution of delta values of digital option obtained from different Monte-Carlo simulations (seeds) with width parameter set to 0.5. 
		}
		\label{fig:DeltaDistribution-Width-05}
	\end{center}
	\addtocounter{cpfNumberOfFigures}{1}
\end{figure}

\begin{figure}[hbtp]
	\begin{center}
		\includegraphics[scale=0.7]{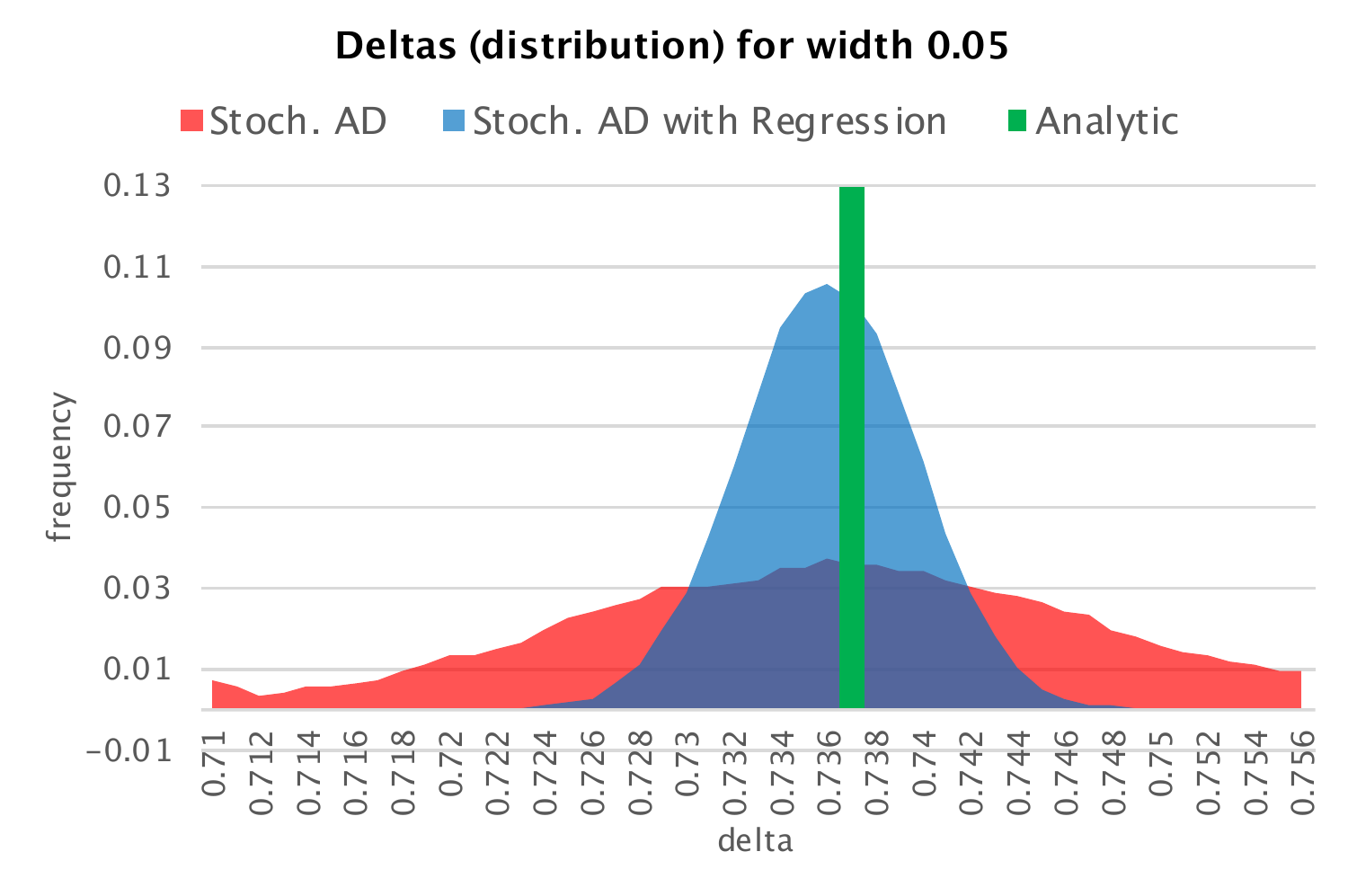}
		\caption[
		]{
			Distribution of delta values of digital option obtained from different Monte-Carlo simulations (seeds) with width parameter set to 0.05. 
		}
		\label{fig:DeltaDistribution-Width-005}
	\end{center}
	\addtocounter{cpfNumberOfFigures}{1}
\end{figure}

	To further analyse the regression method, we compare the two AAD methods for the width parameter $w = 0.5$ and $w = 0.05$. This comparison reveals the advantage of the regression method:
	While AAD with discretized delta function has a much larger Monte-Carlo variance for small width parameters ($w = 0.05$, Figure~\ref{fig:DeltaDistribution-Width-005}), the expectation becomes biased for large width parameters ($w = 0.5$, Figure~\ref{fig:DeltaDistribution-Width-05}). The reason for this effect is that the estimate of the speed of the discontinuity is biased, while the estimate for the density is erroneous. The regression method can correct this issue, by giving an almost unbiased mean \textit{with} an even smaller Monte-Carlo variance.

	The algorithmic differentiation with density regression shows a bias in the expectation, while the method without regression does not exhibit this bias, see Table~\ref{tbl:DeltaDistribution-Width-005}. This might raise a concern. That bias comes from the systematic bias of the regression with the fixed number of paths (20000) determining a fixed size of the regression interval, which results in a fixed second order approximation error in the regression. This error does not vanish if the Monte-Carlo experiment is repeated, but it will vanish if the individual number of Monte-Carlo paths is increased and/or the regression interval is shrunk, see Table~\ref{tbl:DeltaDistribution-Width-0025-reg025}. But even for the configuration in Table~\ref{tbl:DeltaDistribution-Width-005} the bias is well below the Monte-Carlo variance of the algorithmic differentiation with regression (factor 1.8) and far below Monte-Carlo variance of the classic algorithmic differentiation (factor $>6$), such that it is not relevant in a single Monte-Carlo experiment.

	\medskip
	
\begin{figure}[hbtp]
	\begin{center}
		\includegraphics[scale=0.7]{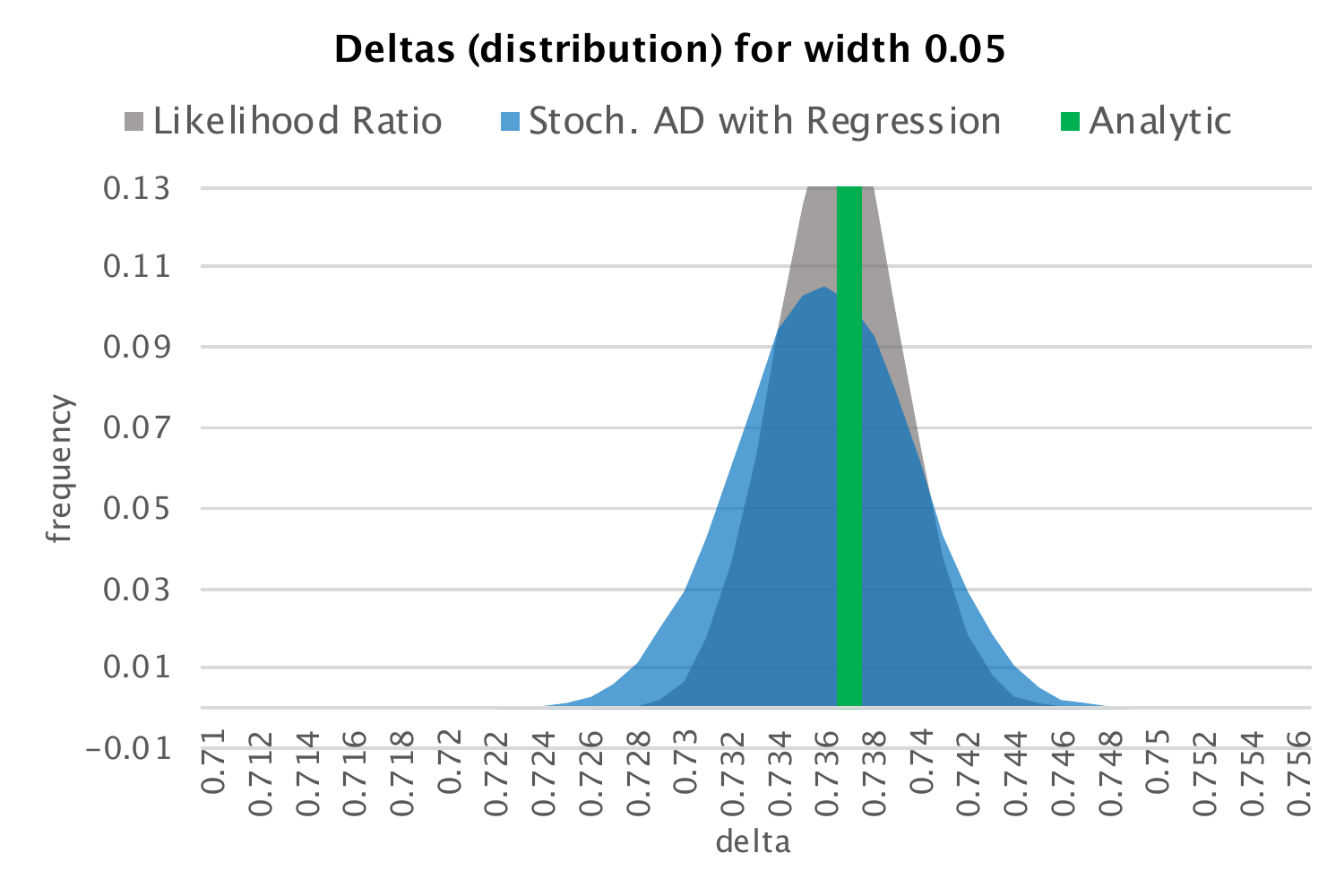}
		\caption[
		]{
			Distribution of delta values of digital option obtained from different Monte-Carlo simulations (seeds): comparing stochastic algorithmic differentiation with regression and the likelihood ratio method. 
		}
		\label{fig:DeltaDistribution-Width-005-aadreg-lr}
	\end{center}
	\addtocounter{cpfNumberOfFigures}{1}
\end{figure}

	Finally, we compare our method, the stochastic algorithmic differentiation with regression, to the likelihood ratio method. Despite its short comings in other situations, the likelihood ratio method is the best numerical method for our test case of a digital option, given that we do not consider variance reduction methods which require modifications of the simulated paths. Figure~\ref{fig:DeltaDistribution-Width-005-aadreg-lr} shows that the stochastic algorithmic differentiation with regression is almost as good as the likelihood ratio method.


%



\begin{table}[p]
	\centering
	\begin{tabular}{ | l || r | r | r | l |}
		\hline
		\multicolumn{5}{|l|}{\cellcolor[gray]{0.95} \vphantom{\Big|} Comparison using width parameter $w=0.5$.}\\ %
\textbf{Method} & Value & \textbf{Bias} & \textbf{Std.~Dev.} & \textbf{Improve} \\ \hline
\hline
Finite Difference & 0.7281 & -0.0079 & 0.0032 &  \\ \hline
Stoch.~AD & 0.7288 & -0.0072 & 0.0033 & 0.97 \\ \hline
Stoch.~AD with Regression & 0.7291 & -0.0069 & 0.0036 & 0.90 \\ \hline
Likelihood Ratio & 0.7361 & 0.0000 & 0.0025 & 1.42 \\ \hline
Analytic & 0.7361 & 0.0000 & 0.0000 & $\infty$ \\ \hline
	\end{tabular}
	\caption{Results from 10000 Monte-Carlo experiments, each using 200000 paths, with width parameter $w=0.5$: While both AAD methods achieve similar variance reductions, the classical approach (corresponding to a pay-off smoothing) has a biased mean.}
	\label{tbl:DeltaDistribution-Width-05}
	\addtocounter{cpfNumberOfTables}{1}
\end{table}

\begin{table}
	\centering
	\begin{tabular}{ | l || r | r | r | l |}
		\hline
		\multicolumn{5}{|l|}{\cellcolor[gray]{0.95} \vphantom{\Big|} Comparison using width parameter $w=0.05$.}\\ \hline\hline
\textbf{Method} & Value & \textbf{Bias} & \textbf{Std.~Dev.} & \textbf{Improve} \\ \hline
\hline
Finite Difference & 0.7359 & -0.0002 & 0.0110 &  \\ \hline
Stoch.~AD & 0.7359 & -0.0001 & 0.0112 & 0.98 \\ \hline
Stoch.~AD with Regression & 0.7355 & -0.0006 & 0.0036 & 3.12 \\ \hline
Likelihood Ratio & 0.7361 & 0.0000 & 0.0025 & 1.42 \\ \hline
Analytic & 0.7361 & 0.0000 & 0.0000 & $\infty$ \\ \hline
	\end{tabular}
	\caption{Results from 10000 Monte-Carlo experiments, each using 200000 paths and width parameter $w=0.05$: The stochastic AAD with regression brings a 3.1 times variance reduction. The use of a likelihood ratio method just gives an additional 42\% improvement.}
	\label{tbl:DeltaDistribution-Width-005}
	\addtocounter{cpfNumberOfTables}{1}
\end{table}

\begin{table}
	\centering
	\begin{tabular}{ | l || r | r | r | l |}
		\hline
		\multicolumn{5}{|l|}{\cellcolor[gray]{0.95} \vphantom{\Big|} Comparison using width parameter $w=0.025$.}\\ \hline\hline
\textbf{Method} & Value & \textbf{Bias} & \textbf{Std.~Dev.} & \textbf{Improve} \\ \hline
\hline
Finite Difference & 0.7362 & 0.0001 & 0.0156 &  \\ \hline
Stoch.~AD & 0.7362 & 0.0002 & 0.0161 & 0.97 \\ \hline
Stoch.~AD with Regression & 0.7359 & -0.0002 & 0.0054 & 3.01 \\ \hline
Likelihood Ratio & 0.7361 & 0.0000 & 0.0025 & 2.11 \\ \hline
Analytic & 0.7361 & 0.0000 & 0.0000 & $\infty$ \\ \hline
	\end{tabular}
	\caption{Results from 10000 Monte-Carlo experiments, each using 200000 paths. The width parameter for the conditional expectation was $w=0.025$. The regression is performed on the distribution with a width parameter of $w=0.25$. Both parameters are half the size as those used for Table~\ref{tbl:DeltaDistribution-Width-005}. The bias on the AAD with regression vanishes, while the improvement of the variance reduction stays similar.}
	\label{tbl:DeltaDistribution-Width-0025-reg025}
	\addtocounter{cpfNumberOfTables}{1}
\end{table}

\clearpage

\subsubsection{Comparing Regression Methods}

For the estimation of $\phi_{X}(0)$ several different approaches may be used. Here we compare two:
\begin{enumerate}
	\item We may use a (localized) regression with basis functions $1$, $x$ (or higher orders) and apply it to samples of the function $d(x) := \frac{1}{x} \mathbb{Q}\left( X \in [0,x] \right)$ (we call this \textit{regression of the density function}).
	
	\item We may use a (localized) regression with basis functions $x$, $x^{2}$ (or higher orders) and apply it to samples of the function $D(x) := \mathbb{Q}\left( X \in [0,x] \right)$ (we call this \textit{regression of the distribution function}).
\end{enumerate}
The two approaches are similar, but not identical. The regression of the distribution gives slightly better variance reduction, but has a slightly higher bias, see Table~\ref{tbl:DeltaUsingDifferentDensities}.

\begin{table}[h]
	\centering
	\begin{tabular}{ | l || r | r | r |}
		\hline
		\multicolumn{4}{|l|}{\cellcolor[gray]{0.95} \vphantom{\Big|} Comparison using different settings for density estimation.}\\ \hline\hline
		\textbf{Method} & Value & \textbf{Bias} & \textbf{Std.~Dev.} \\ \hline
\hline
Finite Difference (0.05) & 0.7359 & -0.0002 & 0.0110  \\ \hline
Stoch.~AD (0.05) & 0.7359 & -0.0001 & 0.0112 \\ \hline
Regression on density (0.05, 0.75) & 0.7352 & -0.0008 & 0.0036 \\ \hline
Regression of density (0.05,0.5) & 0.7357 & -0.0004 & 0.0046 \\ \hline
Regression on distribution (0.05, 0.75) & 0.7346 & -0.0014 & 0.0028 \\ \hline
Regression of distribution (0.05, 0.5) & 0.7355 & -0.0006 & 0.0036 \\ \hline
Likelihood Ratio & 0.7361 & 0.0000 & 0.0025 \\ \hline
Analytic & 0.7361 & 0.0000 & 0.0000 \\ \hline
	\end{tabular}
\caption{Results from 10000 Monte-Carlo experiments, each using 200000 paths. The width parameter for the conditional expectation was $w=0.05$ in all tests. We compare different settings for the estimation of the density. The first value in the bracket gives the value of $w$, the second given the value for the width of the regression interval $w_{\phi}$, where applicable.}
\label{tbl:DeltaUsingDifferentDensities}
\addtocounter{cpfNumberOfTables}{1}
\end{table}

\section{Conclusion}

In this paper we presented a modification of the expected stochastic algorithmic differentiation with a special treatment for the derivative of the indicator function. Our method only relies on the possibility to numerically inspect the random trigger $X$ constituting the argument of the indicator function.

Based on $X$ we can derive an accurate approximation of the density $\phi_{X}(0)$ and define an approximation for the derivative of the indicator function.

The method is an improvement compared to classical path-wise methods and is almost as good as the corresponding (optimal) dual method.

Since the method does not require a modification of the Monte-Carlo simulation (compared to a partial proxy scheme or an importance sampling), the method is computationally efficient. It is generic in the sense that we do not require knowledge about the underlying model generating state variables (e.g., $X$).

With respect to the implementation, the approximation of the derivative of the indicator function can be used in an algorithmic differentiation, as long as the AD algorithm is defined on random variables. We provide a corresponding benchmark implementation. All numerical results presented here can be reproduces by the code in~\cite{finmath-lib}.

\newpage

\printbibliography

\newpage

\section*{Notes}

\subsection*{Suggested Citation}

\begin{itemize}
	\item[] \sloppypar \textsc{Fries, Christian P.}: Stochastic Algorithmic Differentiation of (Expectations of) Discontinuous Functions (Indicator Functions). (November, 2018). \url{https://ssrn.com/abstract=3282667}
	\newline
	\url{http://www.christian-fries.de/finmath/stochasticautodiff}
\end{itemize}

\subsection*{Classification}

\begin{small}
	
	\noindent Classification:
	\href{http://www.ams.org/msc/}{MSC-class}: 65C05 (Primary), 65D25, 68U20.
	\\
	\phantom{Classification:}
	\href{http://www.acm.org/class/1998/ccs98.html}{ACM-class 1998}: G.1.4: Quadrature and Numerical Differentiation.
	\\
	\phantom{Classification:}
	\href{https://www.acm.org/publications/class-2012}{ACM-class 2012}: 10003748: Automatic differentiation.
	\\
	\phantom{Classification:}
	\href{http://www.aeaweb.org/journal/jel_class_system.html}{JEL-class}: C15, G13, C63.
	\\
	

%
	
	\noindent Keywords:
	Algorithmic Differentiation,
	Automatic Differentiation,
	\\ \phantom{Keywords:\ }
	Adjoint Automatic Differentiation,
	Monte Carlo Simulation,
	\\ \phantom{Keywords:\ }
	Indicator Function,
	Object Oriented Implementation,
	\\ \phantom{Keywords:\ }
	Variance Reduction
	
	
	
	\vfill
	
	\bigskip
	\centerline{\small\thepage \ pages. \thecpfNumberOfFigures \ figures. \thecpfNumberOfTables \ tables.}
	
\end{small}

\end{document}